\documentclass[10pt,preprint]{aastex}

\begin{document}

\title{The Vela Pulsar in the Ultraviolet\footnotemark[1]
\footnotetext[1]{Based on observations made with the NASA/ESA Hubble
Space Telescope, obtained at the Space Telescope Science Institute, 
which is operated by the Association of Universities for Research in 
Astronomy, Inc., under NASA contract NAS 5-26555.}}

\author{Roger W. Romani}
\affil{Department of Physics, Stanford University, Stanford, CA 94305-4060}
\author{Oleg Kargaltsev \& George G. Pavlov} 
\affil{Pennsylvania State University, Department of Astronomy \& Astrophysics,
525 Davey Lab, University Park PA 16802}

\email{rwr@astro.stanford.edu, oyk100@psu.edu, pavlov@astro.psu.edu}

\begin{abstract}

	We describe time-tagged observations of the Vela pulsar 
(PSR B0833$-$45) with the {\it HST} {\it STIS} MAMA-UV detectors. Using an optimal
extraction technique, we obtain a NUV light curve and crude FUV phase-resolved
spectra. The pulse is dominated by complex non-thermal emission at
all phases. We obtain approximate spectral indices for the brightest
pulse components, relate these to pulsations observed in other wavebands,
and constrain the thermal surface emission from the neutron star. An upper limit
on the Rayleigh-Jeans component at UV pulse minimum suggests a rather
low mean surface temperature $T_{\rm eff}^\infty <4.6\times 10^{5} (d_{300}/R_{14.2})^2$K.
If confirmed by improved phase-resolved spectroscopy, this observation has
significant impact on our understanding of neutron star interiors.
\end{abstract}

\section{Introduction}

	Young spin-powered pulsars show highly pulsed emission from the radio
to $\gamma$-ray band, arising from narrow acceleration zones in their
active magnetospheres. In the UV to soft X-ray band, however, thermal emission
from the surface can contribute significantly for objects aged 
$\sim 10^4-10^6$yr. Detailed {\it phase-resolved} spectra can isolate these
two components, allowing a measure of the surface spectrum and thermal 
luminosity. Such measurements can constrain the surface composition and,
by measuring thermal emission as a function of age, can probe the equation
of state of matter at supernuclear densities in the neutron star core.
{\it Chandra X-ray Observatory} and {\it XMM-Newton} measurements have 
begun to reveal much about the thermal
spectrum. However, since typical effective temperatures are $kT_{\rm eff} \sim
30-100$\,eV, and interstellar absorption severely attenuates the flux
below $\sim 0.1$\,keV, the X-ray observations of these stars lie well out 
on the Wien tail of a surface thermal spectrum. Two issues then complicate the
interpretation. First, surface composition can dramatically affect the X-ray
flux \citep{r87, zp02} with a light element surface leading
to a Kramers law atmosphere and a large Wien excess. Second, any surface
temperature inhomogeneities will also complicate the spectrum, with hot spots
disproportionately important in the high energy (X-ray) tail.

	For these reasons comparison of the X-ray results with UV 
emission from the Rayleigh-Jeans side of the thermal bump is particularly 
valuable. The challenge here is that non-thermal magnetospheric emission
becomes increasingly dominant as one moves to the red. Fortunately, the 
STIS NUV and FUV MAMA cameras on HST offer access to the UV emission and, 
in TIME-TAG mode, provide the crucial phase-resolved
measurements that allow separation of the thermal and non-thermal fluxes.
We report here on HST STIS observations of the $\tau= 1.1 \times 10^4$yr
Vela pulsar; in a companion paper \citep{kaet05} we report on similar
measurements of the Geminga pulsar ($\tau= 3.2\times 10^5$yr). These
are among the brightest objects spanning the critical age range. Our
observations are complicated by a strong and varying instrumental
background, but we have obtained pulse profiles and crude spectral
information for both objects in the $\sim 5-11$eV energy range.
We compare the pulse minimum thermal fluxes with the X-ray results 
and relate the complex pulsed emission to the multiwavelength
profiles of the Vela pulsar.

\section{HST observations}

	The Vela pulsar (PSR B0833$-$45) was observed on 2002 May 28 
(52422.42-52422.61 MJD) with the Space Telescope Imaging Spectrograph
(STIS). Two orbits were devoted each to Near-Ultraviolet Multi Anode
Micro-channel Array (NUV-MAMA) and Far-Ultraviolet (FUV-MAMA) integrations.
The NUV observations consisted of TIME-TAG mode imaging through
the F25SRF2 filter ($\approx$1400-3270\AA). The useful science exposure time
was 2895s in the first orbit, 3060s in the second.  In the FUV we obtained
low resolution FUV (1140-1730\AA) spectroscopy,
again in TIME-TAG mode, using the 
G140L grating and the 52$\farcs0 \times 0\farcs5$ aperture.
After source acquisition, the exposure time was 2360s and 3015s in the two orbits.
TIME-TAG mode records the (relative) photon arrival time with 125$\mu$s resolution.
The UV detectors have a substantial thermal background (especially
with the increased detector temperatures in these post-SM3a servicing mission
data). For the
FUV data this was partly mitigated by offsetting the target along the slit 
toward the bottom of the array, where a relatively low and uniform thermal 
background allowed more sensitive observations of the pulsar UV continuum.

\subsection{Pulsar Ephemerides}

	To take advantage of the time-tagging, all photon arrival times were
transfered to the barycenter using standard STIS timing routines. A scan
with the STScI-provided routine `checktag.e' revealed only a handful of 
`jump in time' flags in the FUV data; none were found in the NUV data. 
However, we should note that
while the relative event time tags are given to $125 \mu$s, the absolute
time of the event stream is only known to $\sim 10$ms (T. Gull, private
communication). This produces a $\sim 0.1$ absolute phase uncertainty
in the alignment to the radio, X- and $\gamma$-ray pulses.

	To fold the data, we require
a precise pulsar phase. For Vela, this is available from ongoing radio
timing. We used current Parkes ephemerides (courtesy of R.N. Manchester)
from the Australian pulsar timing data file
(http://www.atnf.csiro.au/people/pulsar/psr/archive/data.html). For these
observations $\nu=11.1935036403881$Hz, 
${\dot \nu}= -1.56027\times 10^{-11}{\rm s^{-2}}$, 
${\ddot \nu} = 6.41 \times 10^{-22} {\rm s^{-3}}$ with
barycentric epoch MJD=52408.+$6.7203\times 10^{-7}$.

\subsection{Aperture Extraction of the Pulsar Signal}

Optically, the Vela pulsar is one of the brightest spin-powered pulsars after
the Crab. Its pulsations are in fact directly visible in the MAMA data. 
In Figure 1 we show the central region of the co-added image from our two NUV-MAMA
orbits. The Vela pulsar is at center. No extended structure appears around the
source. An animation of the central field, folded on the pulsar period, is
available at (http://astro.stanford.edu/home/rwr/home.html).

\begin{figure}[t!]
\includegraphics[scale=0.37]{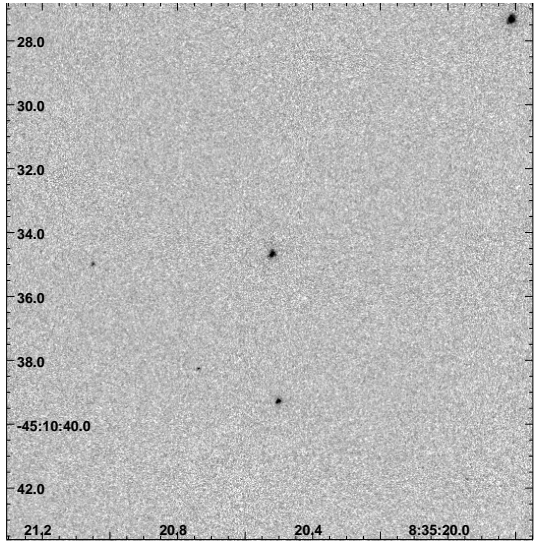}
\caption{NUV-MAMA image with the Vela pulsar at center. 
}
\end{figure}

	A direct 15 pixel (=$0\farcs367$) radius aperture extraction 
of the pulsar gives 5650$\pm$84 counts. The F25SRF2 filter PSF is not tabulated
in the STIS handbook, but measurement shows (Proffitt {et al.} 2002)
that it is nearly identical to that of the F25QTZ filter, whose profile we
obtain from the STIS handbook, Table 14.16. Correcting for the aperture losses
gives a net pulsar signal of 6464$\pm$96
counts or 1.086$\pm$0.016 counts/s. The SRF2 filter has a central wavelength of
2270\AA\, and a 1/2 power range of 1900-3000 \AA\ (with some 
response over 1300-3200 \AA ). Using the NUV MAMA/SRF2 effective 
areas from the STIS handbook Table 14.17, this countrate corresponds to 
$6.18\pm0.09 \times 10^{-18} {\rm erg\,cm^{-2} s^{-1}\AA^{-1}}$ (1.13$\pm$0.02 $\mu$Jy).

	Choice of the best aperture for extraction of the pulsar signal is
somewhat subtle. The large 15 pixel aperture provides a good estimate of
the phase-averaged (DC) pulsar flux, but even this contains only 0.874
of the light of a well-centered PSF. Near pulse minimum, the flux
is only a few percent of maximum. Here the measurement is background limited,
rather than count limited, so a smaller aperture is preferred. Moreover, this
background is time-variable. These factors imply that a time- and phase-dependent 
optimal (PSF-weighted) extraction is needed to obtain the best pulsar S/N, 
especially for the faintest portions of the light curve. So while we adopt 
aperture extraction for phase averaged values, optimal
extraction is used for the time-resolved studies. We describe this extraction 
in \S2.3.

\begin{figure}[h!]
\epsscale{0.5}
\plotone{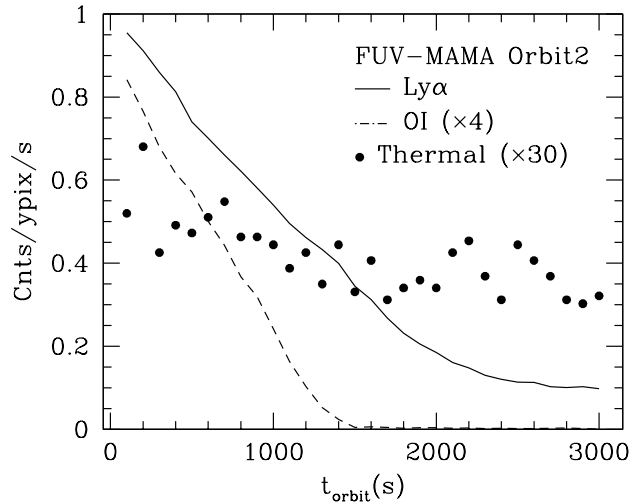}
\caption{FUV-MAMA background rates through the second orbit. The rates are
counts per y (spatial) pixel per second. All fluxes are from a 100 pixel wide
aperture flanking the Vela pulsar. Ly$\alpha$ and OI  fluxes are integrated 
through the line,
while the thermal count rate is measured from a line-free region 
covering spectral pixel range $425<x<800$.
}
\end{figure}

	For the FUV-MAMA spectroscopy, background is even more
important. In particular the strong geocoronal Ly$\alpha$ and OI 1304 lines
dominate the spectrum and vary strongly through the orbit.  The OI] 1356 line 
is detectable, but not strong in these data.  The 0$\farcs$5=20.5pix
slit width projects to 11\AA . We can choose spectral apertures to partly 
avoid these features, however the scattered line flux extends through an 
appreciable portion of the spectrum. Ly$\alpha$ is, for example, the dominant 
background over 1180\AA\ -1245\AA\ .  In addition there is a strong and 
variable background from the thermal glow \citep{land98}. Our observations 
were made post-servicing mission SM3a, and the substantial detector temperatures
caused a large background count rate. We minimized
this background by offsetting the pulsar along the slit to a position with
more modest thermal count rates (y$\sim 105 {\rm pixel} \times 0\farcs0244
{\rm /pixel} =  2\farcs56$ from the bottom of the detector).  
The large (factor $\ga 10$) variations in the geocoronal and thermal backgrounds 
are shown in Figure 2. For an initial
characterization of the FUV spectrum, we selected 16 wavelength bands
avoiding the bright geocoronal lines. Fluxes were obtained from
an 11 pixel wide extraction box, whose centroid followed the expected
y-position of the source. Backgrounds were extracted from 10-pixel wide strips
offset 20 pixels above and below the pulsar. Each photon is weighted by the 
inverse of the normalized flat field $F_{ij}$ at the detector position $(i,j)$.
Summing counts over the spatial aperture and wavelength bin,
the source and background bin counts become $C=\Sigma C_{ij}/F_{ij}$. For
each wavelength bin, we integrate the telescope response $R_\lambda$ (as
a fraction of the unobscured HST aperture $A_{HST}$), the 
slit loss factor $T_\lambda$, the correction for the finite extraction
aperture $H_\lambda$, and the time-dependent sensitivity factor $Q_\lambda$ 
to obtain the corresponding energy flux:
$$
F_\lambda 
= \left ( {{ hc}\over { A_{HST}}}
\right ) { {C_{\rm net}} \over 
{\int R_\lambda T_\lambda H_\lambda Q_\lambda \lambda {\rm d}\lambda}}
$$
These steps follow the standard calibration procedures generally applied to the
phase averaged image as described in the HST STIS data Handbook (\S3),
but allow us to apply corrections to
individual photons, preserving the pulsar phase information for further
analysis. The resulting phase averaged flux spectrum is shown in Figure 3.
The relatively large flux errors are caused by the strong background
corrections that need to be applied to our fixed-width extraction aperture.
Given that this background is highly variable, we will clearly want to go
beyond simple aperture extraction when investigating the phase resolved
spectrum.

\begin{figure}[h!]
\epsscale{0.7}
\plotone{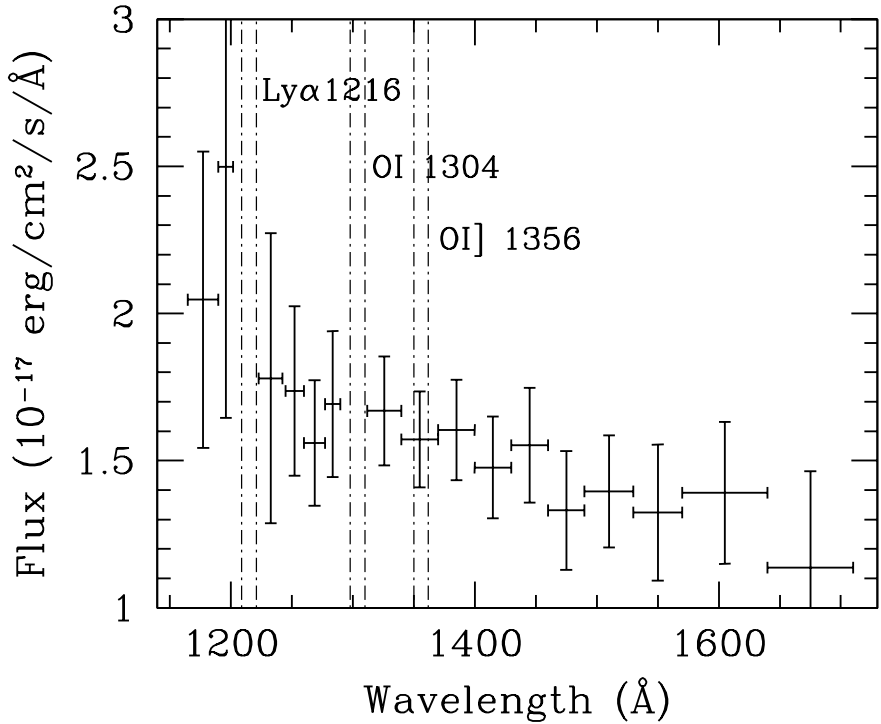}
\caption{FUV-MAMA aperture extracted, phase averaged spectrum.
}
\end{figure}

\subsection{Optimal Extraction of the Pulsar Signal}

	When the response of a detector, the noise properties of the observation
and the detector location of the source PSF are known {\it a priori}, then,
in principle, each detected photon in a given observation time gives
an estimate of the source flux. The HST instruments have relatively
stable and well characterized PSFs, and we can use off-source measurements
to constrain the background; accordingly, we may use a weighted combination of
the observed counts to make the best possible measurements of the
pulsar spectrum, pulse profile and phase-resolved variations.
The classic astronomical approach is optimal extraction (e.g. Naylor 1998).
There are some special features of our application. First, our 
background varies both temporally and spatially (i.e. spectrally).
Second, we need to define a weight for each photon, so that we can
preserve arrival time phase.  Happily, the significant
temporal background variation is on a timescale $\ga 100$\,s, much longer than the 
pulse period. Thus our basic philosophy is to compute an optimal
extraction weight for each position in $100$\,s fiducial time chunks. 
In practice, we compute the weight for each event actually detected in the
extraction window.

	The background is used in two ways in the analysis. First, we need
to subtract the background from the optimally weighted source spectrum.
For this we use windows flanking the source extraction region, with
a total area similar to the central extraction, and measured an average 
background per pixel $B(x,t)$. We also need a statistically
independent estimate of the background and its temporal variation to use 
in computing the optimal extraction
weights. This $B_w(x,t)$, measured from regions flanking the actual
background, needs to be non-zero. As our Poisson data is fairly
sparse, we replace zero backgrounds with the median counts/pixel in the 
line-free region when assembling this weighting factor.

	HST calibration data give an estimate of the normalized spatial PSF
$P_{xy} = P(\delta x,\delta y)$ where 
$\int P(\delta x,\delta y) {\rm d}\delta x {\rm d} \delta y \equiv 1$. 
For the NUV channel, we use a simple radially symmetric
PSF $P(r)$ for the F25QRTZ filter from the STIS handbook (this is
nearly identical to the F25SRF2 PSF; see Proffitt {et al.} 2002).
For the FUV channel the spatial PSF was available for
three different $x$ (wavelengths) for the G140L, which we interpolated
to get $P(x, \delta y)$. Standard extraction provided a good tracing of the pulsar
spectrum central line $y(x)$, so we have $P_{xy}$, normalized at each $x$
along the spectrum. Since we will be treating very low S/N data we will
only be following the spectrum in bins much coarser than the wavelength
resolution, so we do not need to model the spectral extent of the PSF.

	As above, we also can account for the (position dependent)
sensitivity of the detector. For the imaging NUV channel, this is
simply a flatfield correction, $S_{xy}=F_{xy}$. For the FUV spectral data
we additionally include the wavelength dependent effective area
$R_x$, the slit losses $T_x$ and the time-dependent
sensitivity correction $Q_x$ (which is spectrally variable):
$S_{xy}=F_{xy}*R_x*T_x*Q_x$.

	With these definitions, we may `optimally extract' the pulsar
signal from the varying noise. Then we can extract the desired source counts $C$ 
from the data $D_{xy}$:
$$
C = \sum_{xyt} W_{xy}(t) [D_{xy}(t)-B(t)],
$$
where the sum is over some range ${x,y}$ corresponding to the pulsar
source or to a wavelength bin, and the time dependent 
background/pixel $B(t)$ is corrected for the detector response $S_{xy}$
and averaged over the appropriate ${x,y}$ range at time $t$. Here
$$
W_{xy}(t) = P_{xy}S_{xy}[P_{xy}*S_{xy}*C^\prime+B_w(t)]^{-1}N^{-1}
$$
is the source photon weight at $(x,y,t)$ and
$$
N= \sum_{xyt} P^2_{xy}S^2_{xy}[P_{xy}*S_{xy}*C^\prime+B_w(t)]^{-1}
$$
is the weight normalization.

If one has a perfectly known PSF $P_{xy}$, detector sensitivity
correction $S_{xy}$ and  background model
$B(x,y,t)$, then one uses a trial value for the source counts $C^\prime$,
solves for $C$ and iterates to convergence. We find convergence in
$\le 3$ iterations. For the FUV observations the converged phase average
spectrum is in good agreement with (but has smaller errors than) the
aperture extracted spectrum. In the NUV, the 2-D PSF causes greater sensitivity
to PSF shape and centering, and small differences in the average countrate remain
after convergence. Accordingly, we re-normalize the total NUV flux to the aperture
extraction value, but use the optimal extraction weights $W_{xyt}$ to form
the highest S/N light curve.

For example an optimal extraction of the
Vela NUV signal gives a 60-bin light curve with an average $\chi^2$ departure
from constant of 17. The best simple aperture extraction (4 pixel radius)
produces $\langle \chi^2 \rangle$ = 15. Of course, the large (15 pixel radius) 
aperture used for the phase-averaged flux measurement contains too much background
at pulse minimum, giving a rather poor $\langle \chi^2 \rangle \sim 5$. Thus
optimal extraction provides the best light curve S/N with a weighted
combination of the source photons. Because small residual PSF and centering errors
can affect the normalization, we adopt a hybrid approach in our measurement, 
taking the
individual photon optimal weights and applying a small global renormalization
to match the large aperture total counts.


A similar weighting applies for the FUV channel, although the extent of 
the sum $\sum_{xy}$ depends on the desired product. Phase resolved spectra in
coarse wavelength bins will employ modest ranges of $\Delta x$ (which exclude the 
geocoronal emission lines). A simple FUV light curve would employ a summation 
over all $x$, $\delta y$ and $t$, weighting the 
low-noise, low-background time and wavelength slices highly in determining the
weighted flux as a function of pulse phase. Note that to appropriately compute 
weight for counts in the
FUV range, one needs a pre-existing estimate of the FUV pulsar spectrum.

\begin{figure}[t!]
\epsscale{0.7}
\plotone{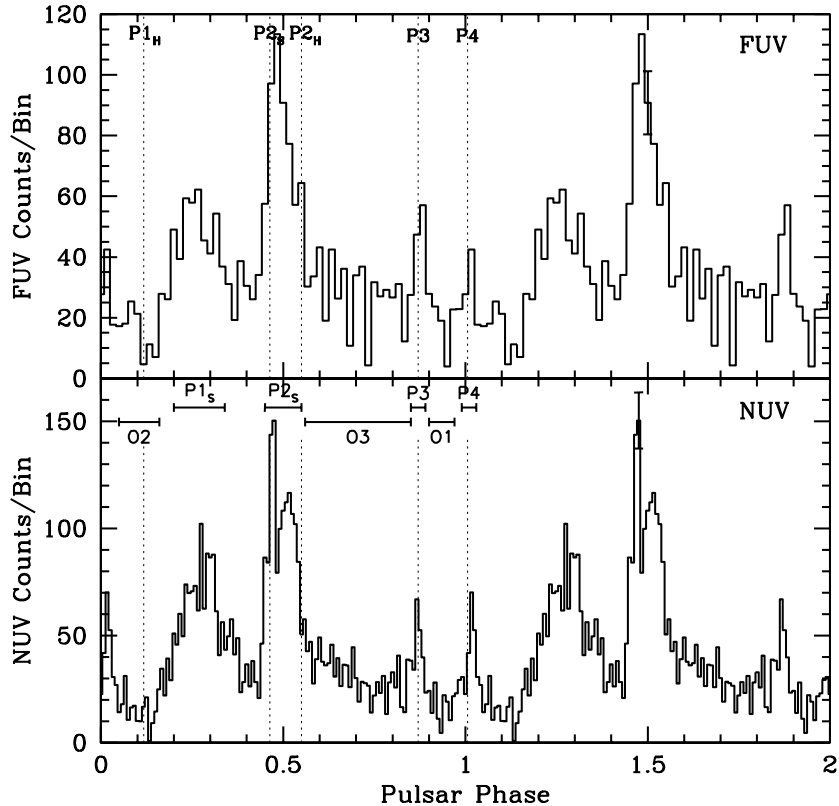}
\caption{NUV and FUV light curves from optimal extractions. Two periods are 
shown for clarity; an example error flag is 
shown in the second cycle. The lack of a flat minimum suggests that non-thermal
emission contributes to all phases.  The dotted lines show the phases
of the 5 pulse components discussed by \citet{ha02}. In the NUV panel
the 4 pulse phase intervals and 3 off-pulse intervals discussed in the text
are shown. The radio peak is at phase $\phi=0$.
}
\end{figure}

\section{Results}

\subsection{Pulsar Light Curve}

	We show the pulse profiles for the two UV channels in Figure 4.
For the NUV data we show a 120 bin (0.74ms/bin) light curve; the noisier
FUV data are plotted with 60 phase bins. Because of the limited absolute
timing information in HST {\it STIS} data, the phases of these
light curves relative to the radio are not independently determined. However
we do see peak structures that correspond well to those seen in the optical
and hard X-ray ({\it RXTE}) bands. Comparing with these, we estimate that the
radio phase 0 is determined within $\sim \delta \phi < 0.02$ (i.e.
$\sim$2\,ms uncertainty).

  Several new features are seen for the first time in these data. First, 
as for the Crab pulsar, there is clearly non-thermal emission (presumably
magnetospheric)
at all phases, since no flat minimum appears in the light curve.
Second, there are at least 4 distinct pulse peaks in the NUV band. Peaks
3 and 4 can be seen in the optical, but were first clearly identified in 
hard X-ray {\it RXTE} data by \citet{ha02}. In our UV data these components 
show much more clearly and, in the higher resolution NUV light curve, we 
see that they are quite narrow. 
In particular P4, the pulse nearest in phase to the radio peak, 
appears unresolved in the NUV data. There is also clear
structure in the main two peaks, seen here for the first time.
The sharp peak in the $P2_S$ component is statistically significant
in the NUV data and has a FWHM no larger than 1ms. This sharp structure is
nearly coincident with, but much narrower than, the soft non-thermal 
component of the X-ray peak (P2$_S$). The broader UV component fills in the
phase between the P2 soft and hard X-ray peaks. The latter peak $P2_H$ is
prominent at GeV energies. The first UV peak, $P1_S$ lies near the optical
peak and is clearly resolved. The UV data suggest that it is bifurcated,
but improved count statistics will be needed to study such structure.
As previously suggested in the optical light curve of \citet{go98},
the pulsed emission trails off following both P1$_S$ and P2$_S$. The
minimum emission is reached during phases $0.05<\phi < 0.15$, our
O2 off pulse interval. Interestingly,
this interval includes the first (brightest) $\gamma$-ray pulse peak.

\subsection{Phase-Resolved Spectra}

	The phase-average spectrum is a rather poor
fit to a power law, with a substantial excess in the shortest wavelength bins.
We should therefore not be surprised that the pulse components show
a range of hardness.
Our FUV statistics do not suffice to provide detailed phase-resolved
spectra of all components of this complex pulse. However, we can examine
the bright peaks in 4 coarse wavelength bins (avoiding the geocoronal lines).
Figure 5 and Table 1 give the NUV+FUV phase-resolved fits to several components.
Here the power law is $F_\nu = F_0 (\nu/\nu_0)^{-\alpha}$, where $F_0$ is the spectral flux
value at $\nu=\nu_0 = 2\times 10^{15}$ Hz.

	In these data we see that while the errors are large, it is clear that
the $P1_S$ and $P2_S$ components are softer than $P3$. The O3 interval,
which includes the tail of P2, is poorly represented by a power law, but
in fact contains the highest 1170 \AA\ count rate, making it relatively hard.
The pulse minimum (O1\&O2 intervals) is in contrast relatively
flat with little FUV excess.

\begin{deluxetable}{llll}
\tabletypesize{\normalsize}
\tablecaption{Fits to Component Spectra [For E(B$-$V)=0.05]}
\tablehead{
Component&$\phi$ & $\alpha$ & Log[$F_0$($\mu$Jy)] 
}
\startdata
$P1_S$& 0.21-0.35&$-$0.37$\pm$0.21 & $-$0.52$\pm$0.03 \\
$P2_S$& 0.46-0.57&$-$0.22$\pm$0.13 & $-$0.43$\pm$0.02 \\
$P3$& 0.86-0.90&\,\,\, 0.45$\pm$0.06 & $-$1.14$\pm$0.01 \\
$P4$& 0.00-0.04&$-$0.30$\pm$0.63 & $-$1.23$\pm$0.08 \\
${\rm O}1$& 0.06-0.174&$-$0.10$\pm0.19^\dagger$ & $-$0.94$\pm$0.02$^\dagger$ \\
${\rm O}2$& 0.91-1.04&$-$0.10$\pm0.19^\dagger$ & $-$0.94$\pm$0.02$^\dagger$ \\
${\rm O}3$& 0.57-0.86&\,\,\, 0.20$\pm$0.39 & $-$0.43$\pm$0.05 \\
\enddata
{\qquad$^\dagger$ Fit for the `Off' = combined O1 and O2 intervals}
\end{deluxetable}

\begin{figure}[t!]
\epsscale{0.7}
\plotone{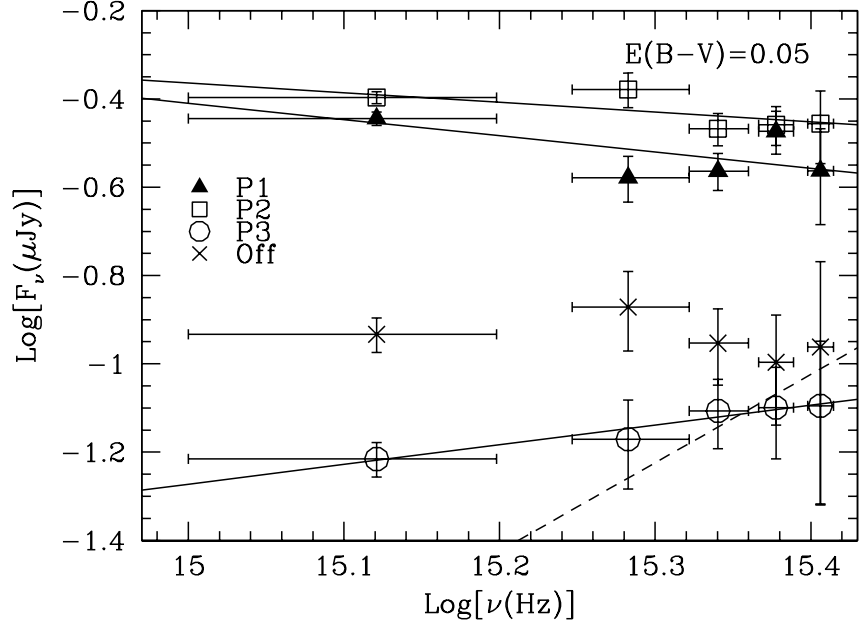}
\caption{NUV/FUV phase-resolved spectra. The observed fluxes are corrected here for
an absorption E(B$-$V)=0.05. Best fit power laws are shown for the three
brightest pulse peaks (solid lines). An upper limit to the Rayleigh-Jeans
($\nu^2$) flux present at pulse minimum is shown with the dashed line.
}
\end{figure}

\section{Discussion and Conclusions}

	Our data have shown that in the UV, the Vela pulse profile is
dominated by multi-component non-thermal emission. To place this pulse
in context, we show in Figure 6 the optical, NUV, hard X-ray and $\gamma$-ray
light curves. In agreement with \citet{ha02} we find that there are at least
five components in the non-thermal pulse profile. The two $\gamma$-ray peaks
($P1_H$ and $P2_H$) are lost below X-ray energies. In the FUV, these
are replaced by $P1_S$ and $P2_S$ which continue to the optical. However,
we find that $P2_S$ is structured, with a softer, narrow `spike' on the
leading edge and a relatively hard component extending into the O3
component. In fact, the soft spike corresponds best to the soft
X-ray component of \citet{ha02}, while our $P2_S$ may contain some of the
hard component, extending into the $O3$ interval.

\begin{figure}[t!]
\epsscale{0.7}
\plotone{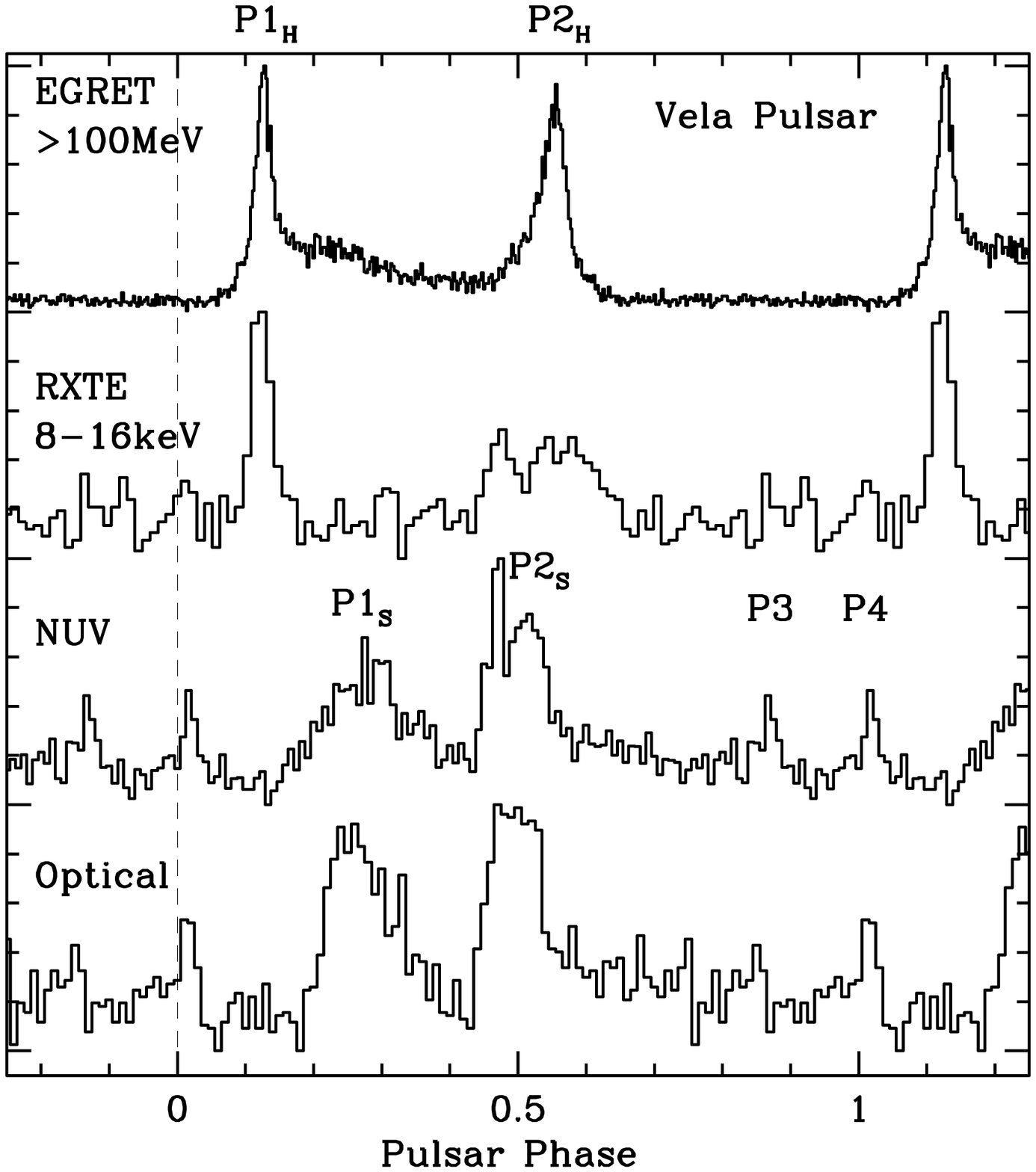}
\caption{EGRET \citep{kan94}, RXTE \citep{ha02}
and optical \citep{go98} light curves plotted with phase 0 set to the 
peak of the radio profile.
}
\end{figure}

	In addition to the narrow spike in $P2$, 
the peaks $P3$ and $P4$ are here seen to 
be very sharp. $P4$ is particularly interesting as it is close to
the radio pulse phase and appears unresolved. For our approximate 
phasing, which is set by
component matching to the optical and {\it RXTE} light curves, the $P4$
component trails the radio maximum by $\sim 1$ms. Interestingly, this is
near the middle of the double peak radio pulse profile measured by
\citet{jet01}, if the intermittent component seen by these authors is
the trailing edge of a hollow cone. This might allow the $P4$ component
to be produced by Compton up-scatter of radio photons from the center of 
the polar cap. However, given the uncertainty of the absolute UV pulse phase,
it is equally plausible to associate this narrow component with the
strong `$\mu$-Giant' pulses \citep{jet01} that lead the main Vela radio 
pulse by $\sim 1.5$ms. This is attractive since true Giant pulse emission
has been associated with narrow non-thermal components at high energy
\citep{rj01}. Given the recent demise of {\it STIS}, the best hope
for testing this idea lies with high statistics optical light curves
with good absolute phasing.

	The physical origin of the various pulse components is
presently unclear. In the outer magnetosphere picture of \citet{r96},
the $\gamma$-ray peaks are argued to arise from caustics generated at
the boundary of the open zone, well above the null charge surface. The
lower energy emission with $P1_S$ and $P2_S$ converging from the
X-ray through optical is then lower altitude emission with the radiation
zone approaching the null charge surface. In the extended slot gap `two pole'
model described in \citet{dhr04} emission extends below the null charge surface
and emission is visible from both poles. This wider range of possible 
emission zones may make it easier to understand the variety of
pulse components seen for Vela. The hard, narrow component $P3$ is perhaps the 
most difficult component to understand, as it disappears at hard X-ray
energies. Although quantitative predictions
are not available, we suggest that this peak may be associated with
inward-directed radiation from the same narrow magnetospheric gaps
producing the outward-going, high energy radiation \citep{crz00}. In
this case only relatively low energy photons can avoid pair production
and traverse the magnetosphere to produce a pulse peak.

\begin{figure}[t!]
\epsscale{0.7}
\plotone{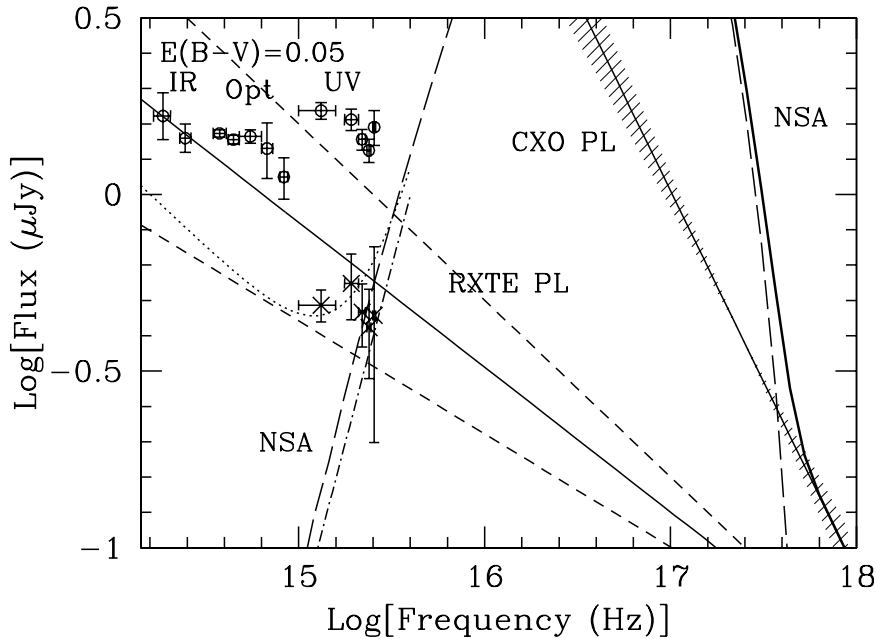}
\caption{Phase-integrated spectrum from the IR through X-ray. We plot our
UV phase-integrated fluxes (error flags include an estimated 5\% flux calibration
uncertainties) along with IR/optical
points from \citet{shi03} and \citet{mica01} (circles). The model plotted is
a H neutron star atmosphere fit to the {\it Chandra} data (NSA, long-dash line) 
along with a power law. Also shown is the range allowed for
a power law fit to the 2.5-20\,keV RXTE phase average spectrum. Our
pulse minimum spectrum is plotted as a DC, unpulsed component (crosses)
along with our upper limit to the Rayleigh-Jeans flux at pulse minimum
(dot-dash line). This line, the DC equivalent of the RJ limit
in Figure 5, lies below the NSA model (long-dashed line).
The curved dotted line shows our upper limit with an added $\nu^{-0.5}$ power law.
}
\end{figure}

	We also wish to compare our spectrum with emission at other
wavelengths. Figure 7 shows the IR-X-ray region of the Vela spectrum,
which is dominated by thermal soft X-ray emission.  Phase-averaged IR and
optical fluxes from \citet{shi03}, 555W, B and U fluxes from
\citet{mica01} and our phase averaged NUV/FUV fluxes, all corrected for
a reddening of E(B$-$V)=0.05, are plotted as open circles. Crudely the
IR-UV data show a flat, $\alpha \approx 0$ spectrum. The ground-based U
and our NUV point seem somewhat discrepant, but we suspect that these
departures are artifacts of imperfect calibration. To compare with the
X-rays, we plot a two component fit to the {\it Chandra} ACIS-S3 + HRC/LETG 
data (thick full line). This model has a magnetic H neutron star atmosphere 
component (NSA, Pavlov et al. 1995) with effective temperature
$T_\infty = 0.68\pm0.01$\,MK and radius $R_\infty=14.2\pm 0.5$\,km at a 
distance of 300\,pc (long dashed line) and a power law component with 
$\alpha = -1.08 \pm 0.08$. A second thermal (i.e. hot polar cap) component
is allowed, but not demanded by the fits.  The soft {\it Chandra} power law clearly 
must break before the UV, as it extends well above the flat IR-UV spectrum.
We also show the $\alpha=-0.41\pm0.09$ power law that best fits the
2.5-20\,keV phase-averaged RXTE data, and its uncertainty range. As
noted by \citet{shi03}, this extends to encompass the optical/IR points,
but again there must be a break to $\alpha \approx 0$ in the UV range.
The individual component spectral indices (Table 1) do not match
the RXTE components measured by \citet{ha02}, but given the evidence for 
a spectral break, the difference in phase structure between the UV
and X-ray and the large errors in the UV spectral indices, this is not 
very surprising. 

	One of our original goals in this observation was to detect or
limit the Rayleigh-Jeans (R-J) portion of Vela's thermal surface spectrum. 
This will of course be most strongly constrained at pulse minimum. 
In Figure 7, we plot as crosses the pulse minimum (O$1$+O$2$) spectrum 
flux when this is an underlying DC (i.e. isotropic) component, after 
correcting for an assumed absorption E(B$-$V)=0.05. Clearly the strongest
constraint on a R-J component comes from the highest frequency FUV points.
Interestingly, these lie below the extrapolation of the NSA model in the
two-component {\it Chandra} fit described above.

	Even at pulse minimum the flux is largely non-thermal and
the maximum R-J contribution depends on what is assumed for this
non-thermal spectrum.
If we assume an $\alpha=-0.5$ power law (PL) contribution, the steepest
power law consistent with the RXTE spectrum, we obtain a limit of
$6.25 \nu_{16}^2\mu$Jy for the R-J component. This is shown as the dot-dash
line in Figure 7, while the combined PL+R-J
spectrum is shown as a curved dotted line. If instead
a flat $\alpha\approx 0$ non-thermal component is assumed,
the maximal R-J contribution is about 20\% smaller.

	In addition to the uncertainties associated with the non-thermal
contribution at this phase, there is also uncertainty associated with
the poorly known extinction. The {\it Chandra} X-ray fit has an absorption column 
$N_H = 2.4\pm 0.2 \times 10^{20}{\rm cm^{-2}}$. With a standard conversion
$N_H \approx 2 \times 10^{21} A_V \approx 6 \times 10^{21} {\rm E(B-V)}$, this gives
an estimate E(B$-$V) =$0.04\pm 0.003$ for Vela's reddening. This is consistent
with line ratios measured for nearby Vela remnant filaments \citep{wb90},
although some filaments show reddening estimates as large as E(B$-$V)=0.1.
For this larger extinction, the phase integrated spectrum has an abrupt jump
in the NUV, with a peak flux of $\sim 2.7\mu$Jy at $\sim 2000$ \AA , a factor
of 1.7$\times$ higher than the IR-optical flux, followed by a decrease in the FUV.
This would be of appreciable interest for interpretation of the non-thermal
emission, but seems less plausible than a simple uniform spectrum.
Thus, our adopted E(B$-$V) = 0.05 represents a reasonable estimate for the 
extinction at the pulsar's position, although larger values cannot be ruled
out. 

	As a final caveat, it is important to
note here that the UV light curve of Geminga measured by \citet{kaet05}
shows an eclipse-like minimum. This minimum was interpreted as the effect of
magnetospheric scattering, which can remove thermal surface flux from
a window of pulse phase near the polar cap. Clearly, if such scattering 
occurs for Vela, the pulse minimum could be suppressed, and we would infer 
an unrealistically low Rayleigh-Jeans temperature. However,
both the O1 and O2 minima have comparable fluxes and seem to
lie near the asymptotic value extrapolated from the O3 window (see Fig. 4).
Thus, a rather special scattering geometry would be needed to suppress the O1/O2
flux without showing up in the light curve. Also, a rather abrupt break in the
non-thermal contribution at $\sim 2 \times 10^{15}$Hz would be needed to 
permit a R-J component comparable to the total flux in the shortest
wavelength FUV bins (see Fig. 5).
Accordingly, we take the $6.25 \nu_{16}^2\mu$Jy limit above as fairly
conservative, and for a $13R_{13}$\,km neutron star at $300d_{300}$pc
we obtain
$$
T_{\rm RJ} \le 3.3 \times 10^5{\rm K} (d_{300}/R_{13})^2 10^{0.4 A_{1210}[{\rm
E(B-V)}-0.05]},
$$
where the last term with $A_{1210}\approx 10$ highlights the dependence 
on the imprecisely known extinction.

At $d=300$\,pc, the {\it Chandra} NSA fit gives $T\, R^2 = 137 {\rm MK\, km^2}$.
However, a light element atmosphere has a Wien excess and, correspondingly, a R-J
depression when compared with the Planck spectrum at the same $T_{\rm eff}$. In fact,
the NSA model above has a flux corresponding to $80 {\rm MK\, km^2}$ in the
R-J regime. This is, however, still 45\% larger than the $55 {\rm MK\, km^2}$
from our pulse minimum flux. Correcting for a rather high extinction 
E(B$-$V)$\ge 0.1$ would make the FUV flux just consistent with the NSA model, 
but at the cost of introducing a peak in the phase-integrated non-thermal 
spectrum, as noted above. There is a precedent for
observed UV flux levels below the extrapolation of the best-fit X-ray thermal 
spectrum.  Our FUV observations of the Geminga pulsar \citep{kaet05} clearly 
detect the $\nu^2$ R-J surface component at $0.45-0.55 R_{13}^{-2}$\,MK;
this is again below the R-J extrapolation of the X-ray fit.  Detailed interpretation
of these UV fluxes will require improved models of the opacity in this band
from low temperature neutron star surfaces. However, we may generally infer that
the R-J emission gives a better indicator of the average surface temperature,
since the X-ray Wien tail can be boosted by unmodeled hot spots and non-thermal
components.

	Thus, with the caveats above in mind, we interpret our measurement as
supporting a very low surface temperature for the young $\tau_c \approx 10^4$yr
Vela pulsar. Even if E(B$-$V) is increased to 0.1 to match the X-ray fit
predictions, the best-fit X-ray surface temperature of 0.68\,MK at R= 14.2\,km is 
already below standard cooling predictions for such a young neutron star 
\citep{tsuet02, yak04}, implying the existence of enhanced neutrino emission, 
e.g. due to kaon core condensates. The non-thermal spectrum then has a peak
in the near-UV. If, on the other hand, our E(B$-$V)$\approx$0.05 estimate of
extinction to the pulsar is supported by further studies, the effective surface
temperature is significantly lower, having important implications 
for the high density equation of state (EOS) and conditions in the neutron star core.
The 0.33\,MK value for blackbody emission from
a canonical 13\,km star would rule out all but the fastest cooling models.
More realistically, if we adopt the $R=14.2$\,km of the X-ray fit and the 
$T_{RJ} = 0.59T_{\rm eff}$ RJ suppression of the NSA model, we infer an 
effective temperature limit $T_{\rm eff} \le 0.46$\,MK. This is quite cool compared to
other young neutron stars.  This temperature would imply that Vela is a high
mass neutron star with rather weak suppression of the URCA process. It would
appear that the softest EOS can now be excluded, since such rapid cooling
favors core pion condensates or even core hyperon matter \citep{yak04}.
Further, models which reach such low temperatures by $10^4$yr for these 
EOS generally do not have strong core proton superfluidity; the minimum 
required mass is $\approx 1.6-1.7M_\odot$, depending on the EOS. Thus,
this low Vela $T_{RJ}$
provides some of the strongest available constraints on matter at high
density. Such an important result certainly argues for further study
and confirming observations.

\acknowledgments

{\small
We are deeply indebted to Divas Sanwal and Slava Zavlin for the help 
with the X-ray data analysis.

Support for program HST-GO-09182 was provided by NASA through a grant from the 
Space Telescope Science Institute, which is operated by the Association 
of Universities for Research in Astronomy, Inc., under NASA contract NAS 
5-26555. Support for this work was also provided by the National Aeronautics and 
Space Administration through Chandra Award Number G03-4078 issued by the 
Chandra X-ray Observatory Center, which is operated by the Smithsonian 
Astrophysical Observatory for and on behalf of the National Aeronautics 
Space Administration under contract NAS8-03060. Additional support for
this program came from NASA grants NAG5-13344 and NAG5-10865.
}

{}


\begin{thebibliography}{}

\bibitem[Cheng, Ruderman \& Zhang(2000)]{crz00}Cheng, K.S., Ruderman, M. \& 
Zhang, L. 2000, \apj, 537, 964

\bibitem[Dyks, Harding \& Rudak(2004)]{dhr04}Dyks, J., Harding, A.K.
\& Rudak, B. 2004, \apj, 606, 1125

\bibitem[Gouiffes(1998)]{go98}Gouiffes, C. 1998, in Neutron stars and Pulsars,
ed. N. Shibazaki, N. Kawai, S. Shibata, \& T. Kifune (Tokyo:Univ. Acad. Press),
363.

\bibitem[Harding {et al.}(2002)]{ha02}Harding, A.K. {et al.} 2002, 
\apj, 576, 376

\bibitem[Johnston {et al.}(2001)]{jet01}Johnston, S. {et al.} 2001,
\apj, 549, L101

\bibitem[Kanbach{ et al.}(1994)]{kan94}Kanbach, G. {et al.} 1994, \astap,
289, 855

\bibitem[Kargaltsev {et al.}(2005)]{kaet05}Kargaltsev, O.Y., Pavlov, G.G., 
Zavlin, V.E. \& Romani, R.W. 2005, 
\apj, in press


\bibitem[Landsman(1998)]{land98}Landsman, W. 1998, Characteristics of the FUV-MAMA
Dark Rate, STIS IDT report

\bibitem[Mignani \& Caraveo(2001)]{mica01}Mignani, R.P \& Caraveo, P.A. 2001,
\astap, 376, 213

\bibitem[Naylor(1998)]{na98}Naylor, T. 1998, \mnras, 296, 339


\bibitem[Pavlov et al.(1995)]{pet95}Pavlov, G. G, Shibanov, Y. A., 
Zavlin, V. E., \&  Meyer, R. D. 1995,
in The Lives of the Neutron Stars, eds. M.A. Alpar, U. Kiziloglu, J. van 
Paradijs (Kluwer: Dordrecht), 71


\bibitem[Proffitt, et al.(2002)]{pro02}Proffitt, C.R., Davies, J.E., Brown,
T.M. \& Mobasher, B. 2002, in 2002 HST Calibration Workshop, S.Arribas,
A. Koekemoer and B. Whitmore, eds., p. 201

\bibitem[Romani(1987)]{r87}Romani, R.W. 1987, \apj, 313, 718

\bibitem[Romani(1996)]{r96}Romani, R.W. 1996, \apj, 470, 469

\bibitem[Romani \& Johnston(2001)]{rj01}Romani, R.W. \& Johnston, S. 2001,
\apj, 557, L93

\bibitem[Sanwal et al.(2002)]{san02}Sanwal, D. {et al.} 2002,
in Neutron Stars in Supernova Remnants, {\it ASP Conf.}, 271, 353

\bibitem[Shibanov et al.(2003)]{shi03}Shibanov, Yu.A., Koptsevich, A.B.,
Sollerman,J. \& Lundquist, P. 2003, \astap, 406, 645

\bibitem[Tsuruta {et al.}(2002)]{tsuet02}Tsuruta, S. {et al.} 2002,
\apj, 571, L143

\bibitem[Wallerstein \& Balick(1990)]{wb90}Wallerstein, G. \& Balick, B. 1990,
\mnras, 245, 701.

\bibitem[Yakovlev \& Pethick(2004)]{yak04}Yakovlev, D.G. \& Pethick, C.J. 2004, 
Ann. Rev. Astron. Astrophys., 42, 169

\bibitem[Zavlin \& Pavlov(2002)]{zp02}Zavlin, V.E. \& Pavlov, G.G. 2002, in
Neutron Stars, Pulsars and Supernova Remnants, Proc. of the 207th Heraeus Seminar,
ed. W. Becker, H. Lesch and J. Tru\"mper (MPE Report 278), astro-ph/0206025

\end{thebibliography}
\end{document}